%% file: main.tex
\documentclass[10pt, conference]{IEEEtran}
\IEEEoverridecommandlockouts
\usepackage{cite}
\usepackage{amsmath,amssymb,amsfonts}
\usepackage{algorithmic}
\usepackage{graphicx}
\usepackage{textcomp}
\usepackage{todonotes}
\usepackage{xcolor}
\usepackage[underline=false]{pgf-umlsd}
\usepackage{tikz}
\usetikzlibrary{shapes,fit,arrows,calc,arrows,arrows.meta,positioning,matrix}
\usepackage{pgfplots}
\usepackage{caption,subcaption}  
\usepgfplotslibrary{groupplots}
\usepgfplotslibrary{fillbetween}
\pgfplotsset{compat=newest}
\usepackage{hyperref}
\usepackage{breakurl}
\usetikzlibrary{pgfplots.groupplots, pgfplots.units}
\pgfplotsset{
	every axis label/.append style={font=\normalsize},
	tick label style={font=\small},
	/pgfplots/enlargelimits=false,
    legend style={legend pos=north east, font=\small},
    legend cell align=left,
    xlabel near ticks,
    ylabel near ticks,
	axis on top,
    highlight/.code args={#1:#2}{
        \fill [every highlight] ({axis cs:#1,0}|-{rel axis cs:0,0}) rectangle ({axis cs:#2,0}|-{rel axis cs:0,1});
    },
    /tikz/every highlight/.style={
        on layer=\pgfkeysvalueof{/pgfplots/highlight layer},
        red!10
    },
    /tikz/highlight style/.style={
        /tikz/every highlight/.append style=#1
    },
    highlight layer/.initial=axis background
}%
\definecolor{maroon}{rgb}{0.5,0,0}
\definecolor{darkgreen}{rgb}{0,0.5,0}
\definecolor{ao}{rgb}{0.0, 0.5, 0.0}
\definecolor{mycolor1}{rgb}{0.0, 0.53, 0.74}%
\definecolor{mycolor2}{rgb}{0.21783,0.72504,0.61926}%
\definecolor{mycolor4}{rgb}{0.93, 0.53, 0.18}%
\definecolor{plum}{rgb}{0.56, 0.27, 0.52}
\definecolor{pinegreen}{rgb}{0.0, 0.47, 0.44}
\definecolor{pthaloblue}{rgb}{0.0, 0.06, 0.54}
\definecolor{saffron}{rgb}{0.96, 0.77, 0.19}
\usepackage{amsmath}
\usetikzlibrary{shapes,fit,arrows.meta,arrows,decorations.markings,calc}
\usetikzlibrary{positioning}
\definecolor{ao}{rgb}{0.0, 0.5, 0.0}

\def\BibTeX{{\rm B\kern-.05em{\sc i\kern-.025em b}\kern-.08em
    T\kern-.1667em\lower.7ex\hbox{E}\kern-.125emX}}

\makeatletter
\newcommand{\linebreakand}{%
  \end{@IEEEauthorhalign}
  \hfill\mbox{}\par
  \mbox{}\hfill\begin{@IEEEauthorhalign}
}
\makeatother

\usepackage{xspace}

\newcommand{\ie}{i.e.,\xspace}
\colorlet{punct}{red!60!black}
\definecolor{background}{HTML}{EEEEEE}
\definecolor{delim}{RGB}{20,105,176}
\colorlet{numb}{magenta!60!black}

\usepackage{listings}
\lstdefinelanguage{json}{
    basicstyle=\scriptsize\ttfamily,
    numbers=left,
    numberstyle=\tiny,
    numbersep=1em,
    xleftmargin=2em,
    showstringspaces=false,
    breaklines=true,
    frame=lines,
    literate=
        {"modelSwaps"}{\bfseries "modelSwaps"}{12}
        {"swapInstance"}{\bfseries "swapInstance"}{14}
        {"stepCondition"}{\bfseries "stepCondition"}{15}
        {"swapCondition"}{\bfseries "swapCondition"}{15}
        {"swapConnections"}{\bfseries "swapConnections"}{17}
        {"modelTransfers"}{\bfseries "modelTransfers"}{16}
}

\begin{document}

\title{fmiSwap: Run-time Swapping of Models for Co-simulation and Digital Twins}



\author{\IEEEauthorblockN{Henrik Ejersbo, Kenneth Lausdahl, Mirgita Frasheri, Lukas Esterle}
\IEEEauthorblockA{\textit{Department of Electrical and Computer Engineering} \\
\textit{DIGIT, Aarhus University}\\
Aarhus, Denmark \\
\{hejersbo, mirgita.frasheri, lukas.esterle\}@ece.au.dk, kenneth@lausdahl.com}
}

\maketitle
\begin{abstract}
Digital Twins represent a new and disruptive technology, where digital replicas of (cyber)-physical systems operate for long periods of time alongside their (cyber)-physical counterparts, with enabled bi-directional communication between them. 
However promising, the development of digital twins is a non-trivial problem, since what can initially be adequate models may become obsolete in time due to wear and tear of the physical components, accumulated errors, or the evolving interaction with the environment. 
As such, there is a clear need for mechanisms that support swapping in new models, as well changing model structures as a whole when necessary.
To address this challenge, we propose in this paper a novel artefact, \textit{fmiSwap}, that is FMI compliant and allows for run-time swapping in standalone co-simulations, where different strategies can be tested easily, as well in fully deployed DT settings with hardware in the loop. 
We adopt a water-tank case-study consisting of a tank and its controller to demonstrate how \textit{fmiSwap} works and how it can support swaps in a safe manner.

\end{abstract}



\input{Sections/intro}

\input{Sections/background}
\input{Sections/maestro-ms}
\input{Sections/experiments}
\input{Sections/conclusions}

\section*{Acknowledgment}
We would also like to thank Jakob Levisen Kvistgaard for his help with the Desktop Robotti case-study.
We would also like to thank Innovation Foundation Denmark and ITEA for funding the UPSIM project. 


\bibliographystyle{IEEEtran}
\bibliography{refs,au}

\end{document}

%% file: Sections/intro.tex
\section{Introduction}\label{sec:intro}



Digital twins (DTs) represent a rather novel approach that consists in building digital replicas of cyber-physical systems (CPSs)~\cite{Fitzgerald&19,Jones2020DT}, \ie systems that combine both software and hardware components potentially over a network.
Such digital replicas are used to follow the behaviour of a CPS during its operation in real-time, in order to make predictions and/or control the CPS if deemed necessary. 
Different components of a CPS can be modelled in a DT depending on the particular needs of a user. 
Once both CPS and DT are deployed in the real world, they will execute parallel to one another, as well as exchange information between them. 
During this time, it is possible that newer -- and better -- models become available, which are able to follow the behaviour of the CPS more accurately. 
Hence, it is desired that these new models replace the old in such a way that the operation of the DT-CPS is uninterrupted. 

In this paper, we deal with precisely such problem, \ie supporting 
upgrades of models in a non-interruptible DT setting.
Specifically, our contribution consists in a fully implemented FMI-based~\cite{FMIStandard2.0.2} mechanism, \textit{fmiSwap}, that allows new or upgraded models to be added runtime in a DT setting. 
Furthermore, it supports upgrading a model structure where the new replacing structure is not known prior to the construction of the initial model. 
This mechanism can further enhance the safety of upgrades. For this, safety parameters need to be defined. Using those, we can perform pre-testing and verification of the effects of dynamic runtime model swapping in a standalone setup. 
This would be decoupled from the physical systems and can be performed before deployment.
We use a water-tank example to show how the proposed model-swap mechanism works in practice. Further details in to the developed mechanism and additional case-studies is presented in~\cite{Ejersbo&23SEAMS}.

The rest of this paper is organised as follows. 
Section~\ref{sec:background} provides a background on FMI and positions our artefact in relation to others.
Section~\ref{sec:maestro-ms} describes in detail the design and implementation of the proposed mechanism, whereas Section~\ref{sec:experiments} reports on a workable experiment with a water-tank example.
Our final remarks are outlined in Section~\ref{sec:conclusions}.


%% file: Sections/background.tex
\section{Background}\label{sec:background}
In this section we cover concepts and tools relevant in the context of this paper: FMI-based co-simulation, and the Maestro co-orchestration engine for co-simulation, as well as position our artefact with respect to similar work in the area.


\subsection{FMI-based Co-simulation}
The design and development of CPSs requires expertise from multiple domains, due to the combination of (networked) software and hardware components. 
These components may depend on different mathematical basis and could be modelled using different tools. 
As individual models/solvers are designed and tested in their corresponding environments, the need arises to test and verify them as a whole. 
A common technique to achieve this goal is co-simulation, 
which couples individual units and executes them coherently in a joint simulation.
This requires a common standard, that describes how units are packaged and interfaced.
In this paper, we adopted the industry standard Functional Mock-up Interface~\cite{FMIStandard2.0.2,modelica.org:Gomes:2021a}. 
Here, simulation units are referred to as Functional Mock-up Units (FMUs). 
FMUs implement a set of c-interfaces (as per the standard), provide a model description file describing their parameters, input, outputs, and are packaged in a specific way. 

A co-simulation can be specified through a multi-model file which describes which FMUs are to be included, the connections between them, and what values to set the parameters to. 
The execution of a co-simulation is carried out by a co-orchestration engine (COE), which implements a co-orchestration algorithm.
The latter describes how the COE can interact with the FMUs, in terms of the order of retrieving outputs and setting inputs, through \texttt{getXXX} and \texttt{setXXX} functions respectively, and progressing them in time through the \texttt{doStep} function. 
Two typical examples of an orchestration algorithm are the Gauss-Seidel and Jacobi.
Both run a loop from a defined start time until a defined end time. 
The former sets the inputs of the first FMU in a co-simulation, requests a \texttt{doStep} such that said FMU progresses from time $h$ to $h+\Delta h$, collects the outputs, and proceeds to the next FMU to be stepped to time $h+\Delta h$.
The latter, sets the inputs for all FMUs, requests a \texttt{doStep} that progresses all involved FMUs to time $h+\Delta h$, collects all outputs, and proceeds to the next time step. 
In this paper we use Maestro2 as our COE, henceforth simply called Maestro, which implements the Jacobian~\cite{Bastian2011} approach.
The Maestro engine consists of the domain specific language called Maestro Base Language (MaBL), an interpreter of the language and utilities to assist in specifying co-simulations in MaBL. 
It is cross-platform, based on the Java, and offers interaction both through a web interface and command line.

\subsection{Novelty of the Artefact}
Model-driven Engineering (MDE) represents a key approach in reducing development complexity for CPSs~\cite{mohamed2021model}, which is a rather cumbersome task involving modelling physical interactions of software systems as well as the interaction among different networked, embedded systems. With MDE, systems are developed using compositions of models. 
As models undergo rapid development, new versions and variants become available with time. 
The challenge lies in swapping these models in a safe way, and verifying that the system behaves in a correct manner after swap.
Replacing or swapping models in MDE, after systems have been deployed, is not entirely new. 
The work on models{@}run.time~\cite{blair2009models} has introduced the idea of replacing individual models while the software system is executed. This allows for runtime adaptation of the behaviour of the respective system. The need for explicit mechanisms and approaches to implement this runtime change and related challenges are highlighted by Bennaceur et al.~\cite{bennaceur2014mechanisms}, G\"otz et al.~\cite{gotz2015adaptive}, Bellman et al.~\cite{bellman2021self}, and Bertolino and Inverardi~\cite{bertolino2019changing} among others.
The recent and very extensive survey on models{@}run.time~\cite{bencomo2019models} highlights the challenge of replacing models during runtime.
In this line of work, Nilsson and Giorgidze~\cite{nilsson2010exploiting} proposed functional hybrid modelling as an approach to non-causal modelling to switch between pre-defined models.
Whereas, Heinzemann et al.~\cite{heinzemann2019transactional} present a modeling language which allows to specify platform-independent models of hierarchical re-configurations. Their reconfiguration protocol guarantees atomicity, consistency, and isolation properties and real-time constraints by design. Furthermore, it can be utilised to verify properties in discrete and continuous physical environments.
In the SEAMS community there are at the time of writing thirty artefacts and model problems, from which three are somewhat conceptually related to what we have proposed in this paper. 
The Hogna platform~\cite{barna2015hogna} enables researchers to quickly deploy and evaluate their self-adaptive algorithms in the cloud, by replacing a component of interest.
Intelligent Ensembles~\cite{krijt2017intelligent} provides a high-level declarative language that supports the specification of dynamic architectures in terms of group formation, as well as a Java runtime library for executing these specifications. 
mRUBIS~\cite{vogel2018mrubis} allows a user to simulate adaptable software, with access to the runtime model of the architecture, which can be exploited by self-adaptation mechanisms. 
Conversely, in this paper we propose a mechanism that supports swapping not only at the level of the model, but also of the model structures themselves, which can be tested accoridingly standalone in simulation, as well as in fully fleshed DT settings coupled to hardware.
The implementation is based on the second version of FMI, FMI2, a widely used standard in industry.

%% file: Sections/maestro-ms.tex
\section{\textit{fmiSwap}: The Maestro Model Swap Mechanism}\label{sec:maestro-ms}
The \textit{fmiSwap} mechanism provides support in the Maestro co-simulation engine for dynamically replacing FMUs, as well as changing multi-model structures, during a running co-simulation. 
Maestro is well-suited for realizing a co-simulation model swap mechanism for two main reasons.
(I) FMUs have well-defined interfaces and states thus allowing for simple and well-defined specifications of multi-models. 
(II) The Maestro interpreted runtime structure allows for the efficient addition of capabilities to dynamically switch to new interpreter contexts with new multi-model specifications.

In the following we first present a conceptual overview of the developed model swap mechanism. 
Thereafter we describe how the conceptual model is implemented into a functioning extension of the Maestro engine.\footnote{Available at \url{https://github.com/INTO-CPS-Association/maestro/releases/tag/Release\%2F2.3.0} and Maven Central}

\subsection{Conceptual Overview and Requirements}\label{sec:conceptual}
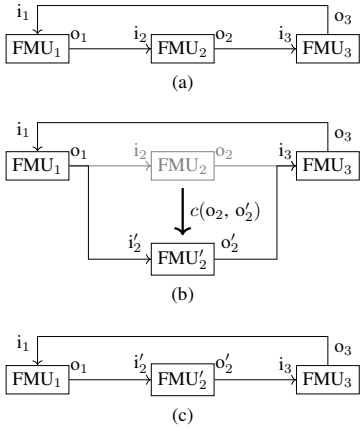
\begin{figure}[tbh!]
    \centering
    \resizebox{!}{0.65\linewidth}{\input{Figures/conceptual}}
    \caption[]{Conceptual overview of the model swap mechanism: (a) Initial scenario with FMUs, input/output variables and connections; (b) Intermediate scenario with both FMU$_2$ and FMU$'_2$ and a swap from FMU$_2$ to FMU$'_2$ based on a condition over variables; (c) Final scenario after the swap.} \label{fig:conceptual}
\end{figure}

A conceptual overview of the mechanism is illustrated in Figure~\ref{fig:conceptual}. 
An existing co-simulation multi-model (MM) is shown in Figure~\ref{fig:conceptual}a, which is transferred at runtime to an intermediate MM containing a new FMU instance FMU$'_2$ in Figure~\ref{fig:conceptual}b, and finally the new instance FMU$'_2$ is replacing FMU$_2$ in the goal MM in Figure~\ref{fig:conceptual}c. 
The replacement is conditionalized by a predicate $c(\mbox{o$_2$, o$'_2$})$ over a subset of MM variables, in this case over outputs $o_2$ and $o_2'$. 
MM (b) may be seen as an extension of scenario (a) where all FMUs of (a) are transferred to (b) with their state preserved. 
MM (c) is the resulting goal scenario after a successful swap. 
In general, the evaluation predicate for a model swap may be more involved than a simple condition over the output variables of the involved FMUs as depicted in Figure~\ref{fig:conceptual}, e.g., an observer FMU could be used that performs more elaborate assessment of outputs.

When specifying a MM for a dynamic model swap, we identify the following required specification parts:
\begin{itemize}
    \item \texttt{swapInstance}: are new instances (FMUs) to be added and eventually swapped for existing ones.
    \item \texttt{swapConnections}: are new connections to and from a swap instance. The new connections of a swap instance are activated (connected) in a conditional manner for both the inputs and the outputs. 
    \item \texttt{stepCondition}: defines the condition for adding a swap instance and connecting its inputs, thus specifies when the instance becomes active with only its inputs connected. 
    This is needed in situations where the initial output of a new instance needs to be synchronized with the output of some other instance before stepping. 
    \item \texttt{swapCondition}: defines the condition for completing the swap connections and connecting the outputs of the new model as specified in the new MM. The new model becomes fully active with connected inputs and outputs, whereas the old is not stepped anymore with its inputs and outputs fully disconnected.
    \item \texttt{modelTransfers}: are existing instances present in the already executing co-simulation that are to be transferred to the new one, unchanged and with all state preserved.
    \item \texttt{parameters} initialization: is required for all newly added instances to ensure a continuous operation when swapping instances, i.e., initial state of a new instance is similar to the final state of the replaced instance (e.g., the initial heading parameter of a vehicle model).
\end{itemize}

In addition to the proposed specification constructs, the internal simulation time of a co-simulation also needs to be transferred to properly continue the execution of a co-simulation. 
This is always required and thus not explicitly stated by our specification constructs.

The FMI specification defines an FMU interface to serialize all state of instances and having this feature generally implemented would ensure that the entire state of an FMU could be serialized and loaded into the context of a new compatible (swap) instance. 
However, such serialization may be difficult to implement, consider as example FMUs with multi-threading models.
Our proposed set of specification constructs can be seen as a simple and general set of specification requirements for FMI-based dynamic model co-simulations, that can handle a broad set of cases avoiding the complexity of full state serialization. 
Our added constructs do not impose new requirements to FMU instances and need handling entirely by the orchestration component. 
Our developed orchestration extensions to the Maestro engine fully support the specified constructs and further allow for dynamic loading of swap specifications to facilitate workflows where new models are dynamically developed and integrated into running simulations. 


\subsection{Specification Format}\label{sec:specformat}
The co-simulation configuration file (multi-model) is in JSON format, with an example given in Figure~\ref{fig:wtspec} (corresponds to the example covered in Section~\ref{sec:experiments}). 
Such file has standard data elements to specify FMU instances and locations (lines 2--7 in Figure~\ref{fig:wtspec}), connections (lines 8--17 in Figure~\ref{fig:wtspec}), and parameters (lines 18--21 in Figure~\ref{fig:wtspec}). 
The model swap mechanism extends the standard format with data elements to specify the proposed new parts, and is given in bold in Figure~\ref{fig:wtspec}.
\begin{figure}[tbh!]
    \centering
    \input{Figures/wtspec}
    \caption[]{Configuration file for the water-tank model swap.}
    \label{fig:wtspec}
\end{figure}

The new configuration elements are specified by entries \verb|modelSwaps| and \verb|modelTransfers|. 
An element in the \verb|modelSwaps| entry has as key an FMU instance to be replaced and as value an object specifying: the FMU instance replacement (\verb|swapInstance|), a conditional expression (\verb|stepCondition|) defining when the FMU replacement can be started (entering the execution phase) and have its inputs enabled, and a conditional expression (\verb|swapCondition|) defining when the FMU replacement output connections will be enabled and the replaced FMU may be unloaded and its connections removed. 
The step condition and the swap condition apply to the input/output swap connections (\verb|swapConnections|). 
In the example MM in Figure~\ref{fig:wtspec}, the new instance is \verb|{x4}.leak_controller|. 
Both input and output connections will become active when both the step and swap conditions evaluate to true, in this case immediately.

\subsection{Specification Interpretation}\label{sec:specinterpret} 
A co-simulation generally consists of three phases: initialization, execution/simulation, and termination \cite{Thule2020}.
The Maestro tool will generate a MaBL specification structured according to these phases - based on the input multi-model configuration file. 
During the initialization phase Maestro first loads and instantiates the specified FMUs. 
Then it calculates the initialization order of the interconnected FMUs based on the topological ordering of FMU port connections - including handling of algebraic loops using fixed point iteration \cite{Thrane2021}. 
A graph is constructed that properly represents dependencies between FMU ports. 
When the multi-model configuration contains model transfers (entry \verb|modelTransfer|), the initialization code in the generated swap specification needs to avoid initializing ports of transferred FMU instances and only consider the initialization of swap FMU instances.
To this end, the proposed mechanism modifies the graph generation to remove edges representing connections into transfer FMUs. 
Note that we check all new fmu instances with respect to any warning signals before attempting to interrupt the simulation.

In the execution phase the simulation loop will be executed. 
This loop iterates the simulation steps of all FMUs and gets and sets input and output variables as described in Section~\ref{sec:background}. 
When the co-simulation contains model swaps (entry \verb|modelSwaps|), the generated simulation loop code needs to reflect the \verb|stepCondition|, \verb|swapConditon|, and \verb|swapConnection| sub-entries. 

The \verb|stepCondition| entry specifies a condition for the FMU swap instance to enter a state where it may be stepped by the COE (state \textit{slaveInitialized}~\cite{FMIStandard2.0.2}), and have its inputs set, thus activated. 
This condition allows a swapped-in FMU to be initialized (enter and leave state \textit{Initialization Mode}) and then started in a controlled manner by a condition over FMU variables - including its own newly initialized variables. 
This is useful when a scenario needs synchronization between newly swapped-in FMUs and already executing FMUs. 
The step condition 
is a Boolean variable initialized to \verb|false|, and evaluated in the simulation loop as a \textit{trigger condition}.
Thus, once it evaluates to \verb|true| it keeps this value. 
The trigger is evaluated at the start of each simulation loop iteration. 
Valid step conditions are expressions build from standard Boolean, relational, and arithmetic operators applied to FMU variables and literal constants as operands. 
A step condition of \verb|true| provides the standard semantics of unguarded entering of state \textit{slaveInitialized} from state \textit{Initialization Mode}. 

The \verb|swapCondition| entry specifies the condition for the FMU swap instance outputs to become effective, thus all input/outputs connections (\verb|swapConnections|) are active by this point (we assume \verb|swapCondition| $\Rightarrow$ \verb|stepCondition|). 
The swap connections affect the code generated to set linked FMU variables (by \verb|setXXX|) . 
For each connected input port of an FMU instance the setting of the port may have to be conditioned by the swap condition. 
The rules controlling this are based on the endpoints of the connection being from swap FMU instances or not. 
E.g.\ for an existing (transferred) FMU instance with an input connected from a swap instance, the setting of that input needs to be guarded by the swap condition of the connected swap instance. 
This because the outputs of that swap instance are only enabled once the swap condition becomes true.
Figure~\ref{fig:mabl} shows an excerpt of the generated MaBL code structure for the example in Figure~\ref{fig:wtspec}. For space considerations, the excerpt has been adapted to show only parts of the auto-generated code. It only shows settings of the \verb|tank.valveControl| input. The \verb|leak_detector| instance has been left out and all error handling has been removed. Comments have been added for readability and are not present in the generated code.
\begin{figure}[tbh!]
    \centering
    \resizebox{!}{.35\textheight}{\input{Figures/mabl_wt.tex}}
    \caption[]{Excerpt of MaBL code with swap and step conditions.}
    \label{fig:mabl}
\end{figure}

To facilitate transfers of values between specifications, our mechanism extends the MaBL language with a new declaration attribute \verb|external|, as used in lines 1--2, to specify a named variable from the context of another MaBL specification, this example the shown FMUs and co-simulation time. 

Lines 17--30 show that the \verb|tank.valveControl| input (valueRef 16) is set to either the \verb|controller.valve| output (line 22) or the \verb|leak_controller.valve| output (line 29) conditionalized by the swap condition. 
Lines 33--37 show that the \verb|controller| instance is only stepped if it has not yet been swapped for the \verb|leak_controller| instance. 
And lines 40--45 show that the \verb|leak_controller| is stepped if the step condition enables it. 
The \verb|tank| instance is always stepped. 
Separate time step variables are used to control the stepping of the transferred instances (\verb|controller| and \verb|tank|) and the swapped-in instance \verb|leak_controller|. 
The transferred instances use a step variable set to the external \verb|START_TIME| transferred from interpreter context. Whereas the swapped-in instance's step variable starts from zero.

\subsection{Execution Overview}
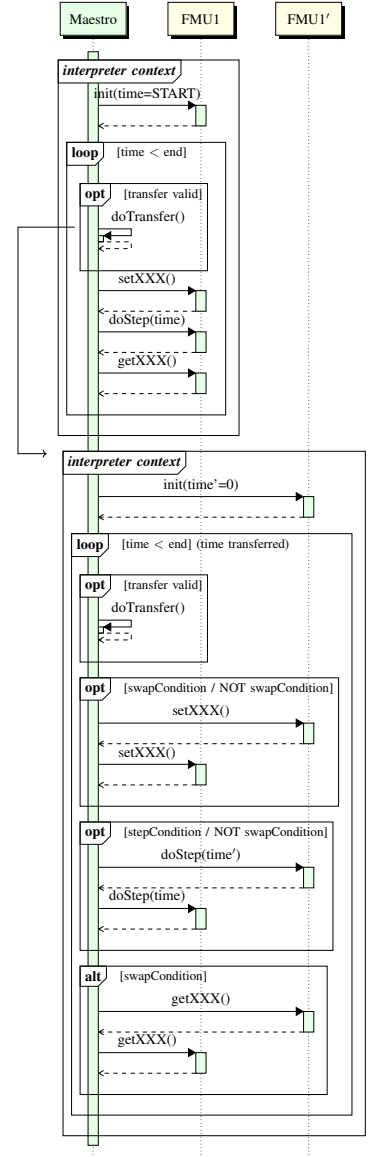
\begin{figure}[tbh!]
    \centering
    \resizebox{!}{.65\textheight}{\input{Figures/swapsequence}}
    \caption[]{Swap specification execution overview.}
    \label{fig:swapsequence}
\end{figure}

The interaction between Maestro and two FMUs, where the simulation starts with FMU1 and swaps-in FMU1$'$, is shown in Figure~\ref{fig:swapsequence}. 
Each simulation loop iteration starts by executing a \textit{transfer point}, where it is checked if a new swap specification is available in a configured file location. 
If so, its validity for transfer is checked. 
If valid, a transfer to a new MaBL interpreter context is performed, where the new specification including the swap-in of FMU1$'$ is interpreted. 
Thereafter, the co-simulation proceeds its execution with the transferred FMU1, and FMU1$'$ loaded from scratch.

Newly loaded FMUs may have separate step conditions, thus the master orchestration algorithm must be able to advance in time individual FMUs accordingly. Thus, in a given global communication step some FMUs may progress (valid step condition) while others don't (invalid step condition).  


%% file: Figures/conceptual.tex
\begin{tikzpicture}[node distance = 1.5cm] 
\node[draw] (FMU2a){FMU$_2$};
\node[draw, right = of FMU2a] (FMU3a){FMU$_3$};
\node[draw, left = of FMU2a] (FMU1a){FMU$_1$};
\node[below = of FMU2a, yshift=40pt] (a) {(a)};

\draw [->] (FMU1a) to node[above, very near start]{o$_1$} node[above, very near end]{i$_2$} (FMU2a);
\draw [->] (FMU2a) to node[above, very near start]{o$_2$} node[above, very near end]{i$_3$} (FMU3a);
\draw [->] (FMU3a.north) -- ++(0pt,15pt) node [above right, very near start]{o$_3$} -| node[above left, very near end] {i$_1$} (FMU1a.north);

\node[draw, below = of a, yshift=15, gray] (FMU2b){FMU$_2$};
\node[draw, right = of FMU2b] (FMU3b){FMU$_3$};
\node[draw, left = of FMU2b] (FMU1b){FMU$_1$};
\node[draw, below = of FMU2b, yshift=10pt] (FMU2pb){FMU$'_2$};
\node[below = of FMU2pb, yshift=40pt] (b) {(b)};

\draw [->, gray] (FMU1b) to node[black, above, very near start]{o$_1$} node[above, very near end]{i$_2$}(FMU2b);
\draw [->, gray] (FMU2b) to node[above, very near start]{o$_2$} node[black, above, very near end]{i$_3$}(FMU3b);
\draw [->] (FMU3b.north) -- ++(0pt,15pt) node [above right, very near start]{o$_3$} -| node[above left, very near end] {i$_1$} (FMU1b.north);
\draw [->] (FMU1b.east) -- ++ (10pt, 0pt) |- node[above, very near end] {i$'_2$} (FMU2pb.west);
\draw [<-] (FMU3b.west) -- ++ (-10pt, 0pt) |- node[above, very near end] {o$'_2$} (FMU2pb.east);

\draw [very thick, -Stealth, shorten >=4pt, shorten <=4pt, ->] (FMU2b.south) to node[right]{$c(\mbox{o$_2$, o$'_2$})$}(FMU2pb.north); 

\node[draw, below = of b, yshift=15] (FMU2c){FMU$'_2$};
\node[draw, right = of FMU2c] (FMU3c){FMU$_3$};
\node[draw, left = of FMU2c] (FMU1c){FMU$_1$};
\node[below = of FMU2c, yshift=40pt] (c) {(c)};

\draw [->] (FMU1c) to node[above, very near start]{o$_1$} node[above, very near end]{i$'_2$} (FMU2c);
\draw [->] (FMU2c) to node[above, very near start]{o$'_2$} node[above, very near end]{i$_3$} (FMU3c);
\draw [->] (FMU3c.north) -- ++(0pt,15pt) node [above right, very near start]{o$_3$} -| node[above left, very near end] {i$_1$} (FMU1c.north);
\end{tikzpicture}

%% file: Figures/wtspec.tex
\begin{lstlisting}[language=json]
{
  "fmus": {
    "{x1}": "watertankcontroller-c.fmu",
    "{x2}": "singlewatertank-20sim.fmu",
    "{x3}": "leak_detector.fmu",
    "{x4}": "leak_controller.fmu"
  },
  "connections": {
    "{x1}.controller.valve": [
      "{x2}.tank.valvecontrol",
      "{x3}.leak_detector.valve"
    ],
    "{x2}.tank.level": [
      "{x1}.controller.level",
      "{x3}.leak_detector.level"
    ]
  },
  "parameters": {
    "{x1}.controller.maxLevel": 2,
    "{x1}.controller.minLevel": 1
  },
  "modelSwaps": {
    "controller": {
      "swapInstance": "leak_controller",
      "stepCondition": "(true)",
      "swapCondition": "(true)",
      "swapConnections": {
        "{x4}.leak_controller.valve": [
          "{x2}.tank.valvecontrol",
          "{x3}.leak_detector.valve"
        ],
        "{x2}.tank.level": [
          "{x4}.leak_controller.level"
        ],
        "{x3}.leak_detector.leak": [
          "{x4}.leak_controller.leak"
        ]
      }
    }
  },
  "modelTransfers": {
    "controller": "controller",
    "tank": "tank"
  }
}

\end{lstlisting}

%% file: Figures/mabl_wt.tex
\lstdefinelanguage{mabl}{
  keywords={typeof, new, true, false, catch, function, return, null, catch, switch, var, if, in, while, do, else, case, break, external},
  keywordstyle=\color{blue}\bfseries,
  ndkeywords={class, export, bool, real, throw, implements, import, this},
  ndkeywordstyle=\color{darkgray}\bfseries,
  identifierstyle=\color{black},
  sensitive=false,
  comment=[l]{//},
  morecomment=[s]{/*}{*/},
  commentstyle=\color{purple}\ttfamily,
  stringstyle=\color{red}\ttfamily,
  morestring=[b]',
  morestring=[b]",
  basicstyle=\scriptsize\ttfamily ,
  numbers=left,
  numberstyle=\tiny,
  numbersep=1em,
  xleftmargin=2em,
}
\begin{lstlisting}[language=mabl]
external FMI2Component controller, tank;
external real START_TIME;

bool swapCondition0 = false;
bool stepCondition0 = false;
real jac_current_communication_point = START_TIME;
real jac_current_communication_point_offset0 = 0.0;

while( ((jac_current_communication_point +
jac_current_step_size) < jac_end_time) )
{
  @Transfer();

  swapCondition0 = swapCondition0 || (true);
  stepCondition0 = stepCondition0 || (true);

  if( (!swapCondition0) )
  {
    // set tank valve input to controller valve output
    tankUintVref[0] = 16;
    tankRealIo[0] = controllerBoolShare[0];
    status = tank.setReal(tankUintVref, 1, tankRealIo);
  }
  if( (swapCondition0) )
  {
    // set tank valve input to leak_controller valve output
    tankUintVref[0] = 16;
    tankRealIo[0] = leak_controllerBoolShare[0];
    status = tank.setReal(tankUintVref, 1, tankRealIo);
  }

  // step controller only if not swapped
  if( (!swapCondition0) )
  {
    controller.doStep(jac_current_communication_point,
    jac_current_step_size, false);
  }

  // step leak_controller if step condition
  if( (stepCondition0) )
  {
    leak_controller.doStep(
      jac_current_communication_point_offset0,
      jac_current_step_size, false);
  }

  // always step tank
  tank.doStep(jac_current_communication_point,
  jac_current_step_size, false);

  // update step times for controller and tank
  jac_current_communication_point =
    jac_current_communication_point
    + jac_current_step_size;

  // update step times for leak_controller
  if( (stepCondition0) )
  {
    jac_current_communication_point_offset0 =
      jac_current_communication_point_offset0
      + jac_current_step_size;
  }
}
\end{lstlisting}

%% file: Figures/swapsequence.tex
\begin{sequencediagram}
\renewcommand\unitfactor{0.5}
\tikzstyle{inststyle}+=[bottom color=green!10, top color=green!10]
\newthread[green!10]{m}{Maestro}
\tikzstyle{inststyle}+=[bottom color=yellow!10, top color=yellow!10]
\newinst[1]{f}{FMU1} 
\newinst[1]{p}{FMU1$'$}

\begin{sdblock}{\textbf{\textit{interpreter context}}}{}
    
\begin{call}{m}{init(time=START)}{f}{} \end{call}

\begin{sdblock}{\textbf{loop}}{[time $<$ end]}
    \begin{sdblock}{\textbf{opt}}{[transfer valid]}
        \begin{call}{m}{doTransfer()}{m}{} \end{call}
    \end{sdblock}
    \node (ctx1) at (current bounding box.west) {};
    \node [yshift=-0.5cm] (cb1) at (current bounding box.east) {};
    \begin{call}{m}{setXXX()}{f}{} \end{call}
    \begin{call}{m}{doStep(time)}{f}{} \end{call}
    \begin{call}{m}{getXXX()}{f}{} \end{call}
\end{sdblock}

\end{sdblock}

\begin{sdblock}{\textbf{\textit{interpreter context}}}{}
    
\begin{call}{m}{init(time'=0)}{p}{} \end{call}

\begin{sdblock}{\textbf{loop}}{[time $<$ end] (time transferred)}
    \begin{sdblock}{\textbf{opt}}{[transfer valid]}
        \begin{call}{m}{doTransfer()}{m}{} \end{call}
    \end{sdblock}
    \node (cb2) at (current bounding box.east) {};
    \begin{sdblock}{\textbf{opt}}{[swapCondition / NOT swapCondition]}
        \begin{call}{m}{setXXX()}{p}{} \end{call}
        \begin{call}{m}{setXXX()}{f}{} \end{call}
    \end{sdblock}
    \begin{sdblock}{\textbf{opt}}{[stepCondition / NOT swapCondition]}
        \begin{call}{m}{doStep(time$'$)}{p}{} \end{call}
        \begin{call}{m}{doStep(time)}{f}{} \end{call}
    \end{sdblock}
    \begin{sdblock}{\textbf{alt}}{[swapCondition]}
        \begin{call}{m}{getXXX()}{p}{} \end{call}
        \begin{call}{m}{getXXX()}{f}{} \end{call}
    \end{sdblock}
\end{sdblock}
\end{sdblock}
\node [yshift=2.8cm](ctx2) at (current bounding box.west) {};
\draw[->] (ctx1) -- ++(-1.5cm, 0pt) |- (ctx2);
\end{sequencediagram}

%% file: Sections/experiments.tex
\section{Water-tank Example}\label{sec:experiments}

In this section, we present a practical example with a water-tank system that shows how to use the proposed \textit{fmiSwap}~\footnote{Instructions on how to use the artefact and reproduce the shown results are given at \url{https://github.com/lausdahl/SEAMS2023Artefact-fmiSWAP}.}.
The water-tank experiment is based on a co-simulation example~\cite{intocps:examples:Mansfield2017} consisting of two FMUs modelling the behaviour of a water-tank and its controller. 
The controller FMU takes as input the current water level in the tank 
and outputs an open or close signal to the tank FMU drain valve in order to keep the water level within configured min $l_{min}$ and max $l_{max}$ levels. 
Meanwhile, water is continuously being added to the tank.

\begin{figure}[tbh!]
	\centering
	\resizebox{\linewidth}{!}{\input{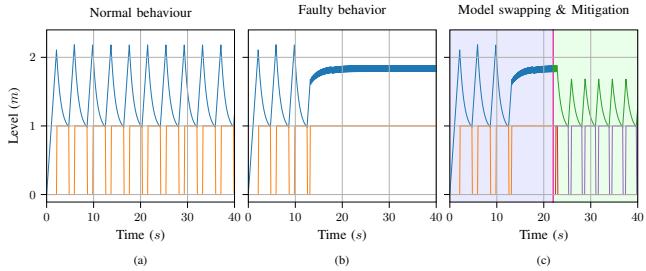}}
	\caption[]{(a) Water level (blue line) and valve control (orange line) of the water-tank in normal behaviour, (b) faulty (fault-injected) behaviour, (c) model swapping to water-tank controller with max level adjustment. Blue/green shaded areas indicate respectively the executions of the old/new models.}
	\label{fig:wtspecres}
\end{figure}

We extend this experiment by using our swap mechanism to dynamically change the model structure by adding a water leak detector and a replacement for the controller that mitigates a detected water leak.
To simulate a faulty water-tank that leaks water, we use a fault injection mechanism~\cite{Frasheri&2021} already developed for the Maestro framework, that allows us to tamper with the inputs and outputs of FMUs. 
In our experiment, we fault inject the input drain valve control of the water-tank FMU by alternating the opening and closing state of the valve in consecutive time-steps. 
The fault occurs at a specified water level lower than the max level of the controller, resulting in a faulty behaviour where the max level is never reached.


During a normal run (Figure~\ref{fig:combo-wt1}) the water level is maintained within $l_{min}$ and $l_{max}$, whereas in the faulty case (configured for a water level of 1.6) it will fall and raise continuously, thus simulating the intended leak (Figure~\ref{fig:combo-wt2})
The top (blue) graph shows the water level of the tank, and the bottom (orange) graph shows the drain valve control (1=open, 0=closed). 
In Figure~\ref{fig:combo-wt2}, the valve control graph does not show the faulty behaviour, due to the fact that the injected values are currently not provided to the logging mechanism of Maestro. 

In a real DT setting, an updated controller could be swapped in that attempts to mitigate the amount of leaked water until the tank could be properly repaired. 
In order to create the conditions for such a swap in this paper we have constructed two new additional FMUs. 
(I) A simple \textit{leak detector} FMU that observes the valve control output of the controller and the water level output of the water-tank. 
If the valve is closed and the water level decreases in three (arbitrary value picked for simplicity) consecutive time steps, the detector FMU will set a Boolean \textit{leak} output to true.   
(II) A simple \textit{mitigating controller} FMU that observes the leak output of the leak detector as well as the current water level in the water-tank. 
If the leak value evaluates to true, the controller will decrease $l_{max}$ by a defined constant before setting its output valve control as usual according to the observed water level. 
As such $l_{max}$ will be adjusted below the leak level, thus reducing the leak flow.

The result of the co-simulation with the run-time swap to the mitigating controller is shown in Figure~\ref{fig:combo-wt3}. The co-simulation starts in the normal behaviour and then enters the fault injected behaviour from simulation time step 12s. 
Until time step 22s the simulation follows the faulty behaviour (blue shaded). 
At time step 22s the swap specification of Figure~\ref{fig:wtspec} is made available to Maestro for pick up (magenta vertical line). 
After the swap (green shaded), the leak detector observes the water level decreasing while the valve is closed. 
Hence it sets its \textit{leak} output to true (red) and this causes the new mitigating controller to reduce $l_{max}$ for the tank. 
Thereafter, the water level is kept between $l_{min}$ and the new $l_{max}'$.

Figure~\ref{fig:wtspec} shows the model swap specification for the dynamic model structure change that adds the water leak detector and the mitigating controller. 
When placed in a specified transition folder the Maestro orchestration engine will pick up this specification and check if a transition is possible.
Note that, Maestro can be configured with a minimum number of steps before offering new specifications and the desired frequency of checking for such specifications.
The specification starts with standard elements specifying FMU locations and port connections between existing (non-swap) FMUs. 
The new model swap part consists of lines 26--48. 
Lines 27--28 specify that the \verb|controller| instance is the target of a swap with the replacement instance \verb|leak_controller|. 
Lines 29--30 specify that the step and swap conditions are both true. 
Thus, the input and output connections of the \verb|swapConnections| take immediate effect when the specification is picked up by Maestro. 
The reason for these trivially true conditions is that the controller FMU is a simple reactive component with no state and as a result needs no synchronization conditions before swapping. 
Lines 31--43 specify the new connections of the \verb|leak_controller| FMU and lines 45--48 specify the FMUs to be transferred. 



%% file: Sections/conclusions.tex
\section{Conclusions and Future Work}\label{sec:conclusions}

In this paper we propose \textit{fmiSwap}, an FMI2 compliant mechanism that allows the run-time swapping of models (FMUs) and whole structures of models (multi-models), that can be used in standalone co-simulations, as well as DT settings with hardware in the loop. 
Thereby providing confidence in the upgraded models prior to deployment in a real uninterruptible DT setting. 
This contribution is rather valuable in the domain of DTs, where there is a clear need for model swapping mechanisms that can perform such a swap while the DT is up and running.
The need stems from the fact that DTs may run alongside their CPS for a long time, resulting in a divergence between the real (actual) and modeled behaviour of the CPS.
Such divergence can be due to inaccurate models, where the error may accumulate over time, but also due to the wear and tear of the physical system or the changes in the environment that render what initially were adequate models useless.
In these circumstances, upgrading the old models becomes crucial for maintaining the safe operation of the DTs.
In future work we will look into setting and tuning parameters on new instances from parameters and outputs of existing instances.
Finally, although the implementation has been realised in FMI2, the principle should apply also to FMI3, as such the implementation will be extended for FMI3 as well.